\documentstyle[a4,12pt]{article}
\catcode`\@=11
\@addtoreset{equation}{section}

\def\@normalsize{\@setsize\normalsize{15pt}\xiipt\@xiipt
\abovedisplayskip 14pt plus3pt minus3pt%
\belowdisplayskip \abovedisplayskip
\abovedisplayshortskip  \z@ plus3pt%
\belowdisplayshortskip  7pt plus3.5pt minus0pt}
\def\small{\@setsize\small{13.6pt}\xipt\@xipt
\abovedisplayskip 13pt plus3pt minus3pt%
\belowdisplayskip \abovedisplayskip
\abovedisplayshortskip  \z@ plus3pt%
\belowdisplayshortskip  7pt plus3.5pt minus0pt
\def\@listi{\parsep 4.5pt plus 2pt minus 1pt
            \itemsep \parsep
            \topsep 9pt plus 3pt minus 3pt}}   
\def\underline#1{\relax\ifmmode\@@underline#1\else
        $\@@underline{\hbox{#1}}$\relax\fi}
\@twosidetrue
\relax
\catcode`@=12

\catcode`\@=11

\def\section{\@startsection{section}{1}{\z@}{3.5ex plus 1ex minus
   .2ex}{2.3ex plus .2ex}{\large\bf}}

\def\ps@headings{\def\@oddfoot{}\def\@evenfoot{}
\def\@oddhead{\hbox{}\hfill
        \makebox[.5\textwidth]{\raggedright\ignorespaces --\thepage{}--
        \hfill }}
\def\@evenhead{\@oddhead}
\def\subsectionmark##1{\markboth{##1}{}}
}

\ps@headings

\catcode`\@=12
\relax

\relax
        
\def\firstpage#1#2#3#4#5#6{
\begin{document}

\begin{titlepage}
\nopagebreak
\title{\begin{flushright}
      \vspace*{-1.3in}  
        {\normalsize  SUSX-TH-98-007 }\\[-5mm]
   {\normalsize SITP-98-001}\\[-5mm]
     {\normalsize hep-th/9810187 }\\[-5mm]
{\normalsize October 1998}\\[.5cm]
\end{flushright}
\vspace{10mm}
{\large \bf #3}}
\vspace{1cm}
\author{\large #4 \\ #5}
\maketitle
\vskip -1mm
\nopagebreak
\begin{abstract}
{\noindent #6}
\end{abstract}  
\vspace{1cm}
\begin{flushleft}
\rule{16.1cm}{0.2mm}\\[-3mm]
$^{1}${\small E--mail: c.kokorelis@sussex.ac.uk}
$^{2}${Part of this work was carried out,
at the School of Mathematical and Physical Sciences
of Sussex University,
in the context of the authors Ph. Thesis.
This work appeared initially in hep-th on 23rd October 1998.}
\end{flushleft} 
\thispagestyle{empty}
\end{titlepage}}
\newcommand{\dal}{\raisebox{0.085cm}
{\fbox{\rule{0cm}{0.07cm}\,}}}
\newcommand{\bb}{\begin{eqnarray}}
\newcommand{\I}{{\cal I}} 
\newcommand{\al}{\alpha}
\newcommand{\be}{\beta}    
\newcommand{\cm}{Commun.\ Math.\ Phys.~}
\newcommand{\pr}{Phys.\ Rev.\ D~}
\newcommand{\pl}{Phys.\ Lett.\ B~}
\newcommand{\ibar}{\bar{\imath}}
\newcommand{\jbar}{\bar{\jmath}}
\newcommand{\np}{Nucl.\ Phys.\ B~}
\newcommand{\e}{{\rm e}}
\newcommand{\gsi}{\,\raisebox{-0.13cm}{$\stackrel{\textstyle
>}{\textstyle\sim}$}\,}
\newcommand{\lsi}{\,\raisebox{-0.13cm}{$\stackrel{\textstyle
<}{\textstyle\sim}$}\,}
\date{}
\firstpage{95/XX}{3122}
{
Generalized $\mu$-Terms from Orbifolds and M-Theory}
{ Christos Kokorelis$^{1,2}$}
{\normalsize
Sussex Institute for Theoretical Physics\\
\normalsize 11 Hanover Court, Wellington Road\\
\normalsize Brighton,  BN2 3AZ, U. K }    
{We consider solutions to the 
$\mu$-term problem
originating in 
the effective low energy theories, of $N=1$ $Z_N$ orbifold 
compactifications 
of the heterotic string, after supersymmetry breaking. They are 
consistent with the invariance of the one loop corrected effective action 
in the linear representation for the dilaton.
The proposed $\mu$-terms naturally generalize solutions proposed 
previously, in the literature, in the context of minimal low-energy 
supergravity models.
They emanate from the connection of the non-perturbative superpotential
to the determinant of the mass matrix of the chiral compactification
modes.
Within this approach we discuss the lifting of our solutions to their
M-theory compactification counterparts.
}
\newpage

\section{introduction} 

One of the necessary ingredients of the minimal supersymmetric extension
of the standard model of electroweak interactions is the existence in its
matter superpotential of a mixing term between the two Higgs doublets,
namely $W_{tree}= \mu H_1 H_2$.
The coupling of this bilinear term is the so called $\mu$-term.    
Its presence introduces a hierarchy problem in the theory.
Clearly the presence of such a term at the electroweak scalar potential
of the theory
is essential in order to avoid the breaking of the 
Peccei-Quinn
symmetry and the appearance of the unwanted axion and to give 
masses to the d-type quarks and e-type leptons which otherwise will
remain massless.
In addition the presence of the $\mu$-term is necessary for the correct 
electroweak symmetry breaking. However during the latter process
the low energy parameter $\mu$, of the 
electroweak scale, is identified with a parameter of order 
of the Planck scale something unacceptable. 
An explanation of the origin of $\mu$ term generates the 
$\mu$-problem and several
scenaria have appeared in the literature providing a solution.

Mechanisms for the generation of the $\mu$ term
make use of gaugino condensation to induce an effective
$\mu$ term \cite{casa} or the presence of mixing $H_{IJ}$ terms in the
K\"ahler potential \cite{gima,casa,ANTO}, which induce after 
supersymmetry breaking an effective $\mu$ term 
of order ${\cal O}(m_{3/2})$.
In the context of supergravity models coming from string vacua, with 
spontaneusly broken supersymmetries, the first examples of a
$\mu$-term generated by the K\"ahler potential
appeared in \cite{sp1}. A general discussion of the generation of the 
$\mu$ and B-terms in the context of perturbative and 
non-perturbative supersymmetry breaking appeared in \cite{sp2}.     
Another solution, applicable to supergravity models, makes use of
non-renormalizable terms (fourth or higher
order) in the superpotential. They have the form
$M_{Pl}^{1-n}A^{n}H_1 H_2$ and generate a contribution \cite{kini} to
the $\mu$ term of order
$\sim {\cal O}(M_{Pl}^{1-n} M_{hidden}^n)$ after the 
hidden fields A acquire a vacuum expectation value.
Superpotential generation of $\mu$-terms in the context of freely 
acting orbifold
compactifications of the heterotic string on the $K_3 \times T^2$,
appeared in \cite{culu}. The latter was examined in
F-theory compactification on Calabi-Yau 4-folds in \cite{witta}.

In this paper we explore the origin of $\mu$ terms in $(2,2)$ orbifold 
compactifications of the heterotic string.  
We discuss particular solutions to the $\mu$ problem related 
to the generation of the mixing terms 
between Higgs fields behaving as neutral scalar moduli in the K\"ahler 
potential.
They are related to the presence 
of duality symmetries originating from subgroups of the modular group
$PSL(2,Z)$. 
The latter appears in non-decomposable (2,2) symmetric $Z_N$ orbifold
compactifations of the heterotic string in four dimensions
and in certain orbifold limits of $K_3 \times T^2$.
Previous solutions in the literature of generating $\mu$ for the 
$SL(2, Z)$ case appeared in \cite{ANTO,lust}.

Lets us explain the origin of such mixing terms in 
superstring theory. We assume that our effective theory of 
the massless modes after compactification is that of the heterotic 
string preserving $N=1$ supersymmetry while our effective theory
is described by the usual $N=1$ two derivative supergravity.
Let us fix the notation \cite{dkl1} first.
We are labeling the ${\bf 27}$,$\;{\bf \bar 27}$ with letters from the
beginning (middle) of the Greek alphabet while moduli are associated with
latin characters.
The gauge group is $E_{6} \times E_{8}$, the matter fields are
transforming under the ${\bf 27}$,$\;{\bf (\bar 27)}$ representations
of the $E_{6} $, ${\bf 27}$'s are related to the $(1,1)$ moduli while
${\bf (\bar 27)}$'s are related to the $(2,1)$ moduli in the usual 
one to one correspondence. The K\"ahler potential is given by
\begin{equation}
K = B^{\alpha}B^{\bar \alpha} Z_{\alpha{\bar \alpha}}^{(1,1)} + 
C^{\nu} C^{\bar \nu} 
Z_{\nu{\bar \nu}}^{(2,1)} + (B^{\alpha} C^{\nu} H_{\alpha \nu} + c.c) +
\dots
\label{papa}
\end{equation}
with the B and C corresponding to the ${\bf 27}$'s and ${\bf {\bar
27}}$'s respectively.
Appropriate expansion of the low energy supergravity observable scalar 
potential,
after supersymmetry breaking, generates a general contribution
in the form
\begin{equation}
{\mu}_{IJ} B^I C^J,\; {\mu}_{IJ} =   m_{3/2} H_{IJ}   
-{\bar F}^{\bar j}{\bar \partial}_{\bar j} H_{IJ} + {\tilde \mu}_{IJ},\;
H_{IJ}= \frac{1}{(T + {\bar T})(U + {\bar U})},
\label{term}
\end{equation}
where F is the auxiliary field of the $(1,1)$ or $(2,2)$ moduli
and have assumed that matter fields and moduli come
from the same complex plane.
In this work, as the contributions to the $\mu$-term coming
from the first two terms in (\ref{term}) are standard, we will examine  
the origin of the additional $\tilde \mu$-term.
Our study requires an expansion of the superpotential, confirmed 
posteriori\footnote{The solution
to the $\mu$-problem proposed in \cite{casa} in the context of 
a general minimal supergravity,
required the same general expansion of W.}, in the form
\begin{equation}
{\cal W}\;=\;{\cal W}_{o}\; +\; {\cal W}_{BC}\;B C.
\label{dgggin}
\end{equation}
In this case 
the $\mu$ term receives an additional contribution in
the form
\begin{equation}
{\tilde \mu}_{BC} =e^{G/2} {\cal W}_{BC}.
\label{additinal}
\end{equation} 
The superpotential of the theory in the form (\ref{dgggin}) comes
from non-perturbative effects since
terms in this form don't arise in perturbation theory, 
due to
non-renormalization theorems \cite{BD1,dsww}.
Furthermore, because supersymmetry cannot be broken by
any continous parameter\cite{BD1}, the origin of such terms
may not come from a spontaneous breaking version of supersymmetry but
necessarily its origin must be non-perturbative.

In the beginning of section 2, we will exhibit the method of 
calculating superpotentials that receive contributions from the 
integration of massive modes. For this reason we will use
initially the non-decomposable orbifold $Z_6 -II-b$ \cite{deko1}. In this case 
the expansion of the superpotentials into the form
$W\;=\;W_{o}\; +\; W_{BC}\;B C$, is consistent with the invariance
of the one-loop corrected effective action, in the linear representation
of the dilaton,
under tree level ${\Gamma}^{o}(3)_T$ transformations,
\begin{equation}
T \stackrel{\Gamma^o(3)_T}{\rightarrow} \frac{aT-ib}{icT+d},\;c\equiv\;
0\;mod\;3,
\label{aditi}
\end{equation}
which leave
the tree level K\"ahler potential 
\begin{equation}
K= -\log[(T + {\bar T})(U+{\bar U}) -(B+ {\bar C})(C + {\bar B})]
\label{ertklio}
\end{equation}
invariant, only if
$ W \rightarrow (i c T + d)^{-1} W$ and
\begin{eqnarray}
{\cal W}_{o} \rightarrow (i c T + d)^{-1}{\cal W}_{o},\;and&
{\cal W} _{BC} \rightarrow
(i c T + d)^{-1} W_{BC} + i\;c\;W_{o}.
\label{oups}
\end{eqnarray}

The paper is organized as follows. 
In sect. 2 we analyse the
different modular orbits appearing
in the moduli space for the 
non-decomposable orbifolds.
Next, we describe
the identification of the non-perturbative superpotential $\cal W$ with
the mass matrix of the chiral masses of the compactification modes.
We perform the sum over modular orbits, integrating
the $N=2$ massive untwisted states of the compactification.
In this respect, we calculate $\cal W$ in four dimensions by
taking into account contributions from general non-decomposable
$N=1$ $(2,2)$ symmetric Coxeter orbifolds.
In addition, we calculate the dilaton dependence of the non-perturbative
superpotentials either by using the BPS sums or by the use of the
gaugino condensates. We identify the $W^{non-pert}$ for all classes of 
non-decomposable Coxeter orbifolds.
In sect. 3 we identify the relevant orbifolds appearing in the
classification list of \cite{erkl} and which can be characterized as
generalized Coxeter orbifolds  and calculate $\cal W$.
In sect. 4  we describe the contributions to the $\tilde \mu$ term
coming from sections 2 and 3.
As we will see, this analysis allows to
describe a variety of possible phenomenological scenarios.
In sect. 5 we explain the promotion of the soft terms, and
in particular the B soft-term arising through
the $\mu$-terms of section four, to their counterparts coming from
M-theory compactifications to four dimensions.
In sect. 5 we summarize our conclusions.

\section{Non-perturbative Superpotentials from Modular Orbits of\\ 
Coxeter Orbifolds}

Let us consider first the generic case of an orbifold
where the internal torus
factorizes into the orthogonal sum $T_6 = T_2 \oplus T_4$ with the 
$Z_2$ twist acting on the 2-dimensional torus lattice.
We will be interested in the mass formula of the untwisted subspace
associated with the $T_2$ torus lattice.
In this case, the momentum operator factorises into the
orthogonal components of the sublattices with 
$(p_{L};p_{R}) \subset  {\Gamma}_{q+2;2}$ and            
 $(P_{L};P_{R}) \subset  {\Gamma}_{20-q;4}$ . And as a result the 
mass operator $M^2$ factorises as below while the spin
S for the ${\Gamma}_{q+2;2}$
sublattice becomes
\begin{equation}
\frac{\alpha^{\prime}}{2} M^{2}= p_{R}^{2} + P_{R}^{2} +2N_{R},\;p_{L}^{2} -
p_{R}^{2}= 2(N_{R}+1-N_{L})+ \frac{1}{2} P_{R}^{2} -
\frac{1}{2} P_{L}^{2}=2 n^{T} m + b^{T} C b,
\label{arxidi1}
\end{equation}
where C is the Cartran metric operator for the invariant
directions of the sublattice ${\Gamma}_{q}$ of the $\Gamma_{16}$
even self-dual lattice\footnote{Compatibility of the
untwisted moduli with the twist action on the gauge coordinates
comes from the non-trivial action of
the twist in the gauge lattice. This means that that the untwisted
moduli of the orbifold have to the equation M A = A Q, where
Q, A and M represent the internal, Wilson lines and gauge twist
respectively.}.
The above formula involves perturbative BPS states which
preserve $1/2$ of the supersymmetries, which belong to short
multiplet representations of the supersymmetry algebra.

Let us now consider the $Z_6 -II-b$ orbifold. 
 This orbifold is non-decomposable 
in the sense that the action of the lattice twist does not decompose in
the orthogonal sum  
$ T_6 = T_2 \oplus T_4$ with the fixed plane lying in $T_2$. Its complex
twist is $\Theta=(2,1,-3)(2 \pi i)/6$.
The orbifold twists $\Theta^2$ and $\Theta^4$, leave the second  
complex plane unrotated. The lattice in which the twists $\Theta^2$
and $\Theta^4$ act as an lattice automorphism is the $SO(8)$. In addition 
there
is a fixed plane which lies in the $SU(3)$ lattice and is associated with
the $\Theta^3$ twist.
The action of the internal twist can                           
be made to act as $-\;I_2$ on a $T_2$ by appropriate parametrization
of the momentum quantum numbers.
Consider now a k-twisted sector of a six-dimensional orbifold of the
the heterotic string associated with the twist $\theta^{k}$.
If this sector has an invariant complex plane then its
twisted sector quantum numbers have to satisfy 
\begin{eqnarray}
Q^{k}n=n,\;\;Q^{*k}m=m,\;\;M^{k}l=l,
\label{axid1}
\end{eqnarray}
where Q defines the action of the twist on the internal lattice and
M defines the action of the gauge twist on the $E_{8} \times E_{8}$
lattice.
This 
means that with
\begin{eqnarray}
n=\left(\begin{array}{c}{\tilde n}_1 \\{\tilde n}_2
\end{array}\right),\;m= \left(\begin{array}{c}{\tilde m}_1\\
{\tilde m}_2\end{array}\right),\;l= \left(\begin{array}{c}{\tilde l}_1\\
{\tilde l}_2
\end{array}\right)
\label{bre1}
\end{eqnarray}
and with $E_a$, a=1,2 a set of basis vectors in the fixed directions of the 
orbifolds, $\tilde E_{\mu}$, $\mu=1,\dots,d$ the set of basis vectors in
the fixed directions of the gauge lattice,
the momenta take the form
\begin{equation}
P^{\prime}_{L} = ({\rho}\frac{\tilde m}{2}+(G_{\bot}-B_{\bot}-
\frac{1}{4}A_{\bot}^{t}C_{\bot}A_{\bot}){\tilde n}-\frac{1}{2}
A_{\bot}^{t}C_{\bot}{\tilde l},\;{\tilde l}+A_{\bot} 
{\tilde n})
\label{kij2}
\end{equation}
\begin{equation}
P^{\prime}_{R} =({\rho}\frac{\tilde m}{2}\;\;\; +
(G_{\bot}\;\;-\;\;B_{\bot} + \frac{1}{4}A_{\bot}^{t} C_{\bot} 
A_{\bot}){\tilde n} - \;\;\frac{1}{2} A_{\bot}^{t} C_{\bot} 
{\tilde l}, 0 ),
\label{kij1}
\end{equation}
$\rho_{ab}\stackrel{def}{=}E_a \cdot {\tilde E}_{\mu}$.
Here $\rho$ is the $\rho_{ab}$ matrix, $C_{\bot}$ is the Cartran matrix
for the fixed directions and  $A_{\bot}^{Ii}$ is the matrix for the
continous Wilson lines in 
the invariant directions $i=1,2,\;I=1,\dots,d$. $G_{\bot}$ and $B_{\bot}$  
are
$2 \times 2$ matrices and ${\tilde n},\;{\tilde m},\;{\tilde l}$ are the 
quantum numbers in the invariant directions.

For orbifold compactifications, where the underlying internal torus 
does 
not decompose into a $ T_6 = T_2 \oplus T_4 $ , the $ Z_2 $ twist 
associated with the reflection $ - I_2 $ does not put any
additional constraints on the moduli $U$ and $T$. As a consequence
the moduli space of the untwisted subspace is the same as in 
toroidal compactifications.
For the $Z_{6}-II-b$ orbifold,  
${\rho} = \left( \begin{array}{cc}
1&0\\0&3\end{array}\right)$.
The mass formula \cite{deko1} for the $\Theta^2$ subspace reads 
\begin{equation}
m^2 = \sum_{m_1, m_2 \atop n^1,n^2} \frac{2}{Y} 
|-T U'n^2+iTn^1- iU'm_1+3m_2|^2_{U'= U-2i} = {|{\cal M}^{pert}|^2}/{(Y/2)},
\label{kavou} 
\end{equation}
\begin{equation}
Y = (T +{\bar T})(U+{\bar U}).
\label{kavo1} 
\end{equation}
The quantity $Y$ is connected to 
the K\"{a}hler potential, $ K = - \log Y $ .
The target space duality group is $\Gamma^0(3)_T \times 
\Gamma^0(3)_{U'}$ where $U' = U -2i$.
Let us now connect $\cal W$ \cite{fklz} to the 
target space partition function Z coming from the integration of the
massive chiral compactification modes
\begin{eqnarray}
Z=e^{-F_{fermionic}}= - det(({\cal M}^{pert})^{\dagger}{\cal M}^{pert}) =
- \frac{|W|^2}{Y},\;\nonumber\\
F_{fermionic}=\sum_{(n,m) \neq (0,0)} \log det(({\cal M}^{pert})^{\dagger}
{\cal M}^{pert}).
\label{qua71}
\end{eqnarray}
Here ${\cal M}$ represents the fermionic mass matrix, 
F the topological free energy and $\cal W$ the perturbative
superpotential.
Working in this way, we define the free energy as the one
coming from the 
integration of the massive compactification modes, i.e. 
Kaluza-Klein and
winding modes. Because non-compactification modes like
massive oscillator modes are excluded from the sum
the free energy can be characterized as topological\cite{fklz,oova}.
From (\ref{qua71}) we can deduce that
\begin{equation}
W = det {\cal M}^{pert}.
\label{kiopiasro}
\end{equation}
Here ${\cal M}^{pert}$ is the mass matrix of the chiral hypermultiplet masses 
of the massive compactification modes.
Note that in order to identify (\ref{kiopiasro}) as a non-perturbative 
superpotential in order to exhibit an $e^{-S}$ dependence, we will connect
M to special geometry of a vector multiplet sector. 
According to the latter the N=2 non-perturbative mass matrix is identified
with the BPS short multiplet expression
\begin{equation}
{\cal M}^{non-pert}_{BPS}= 2e^K {\cal M}= {2}e^{K}|M_I X^I + iN_I F^I|^2,
\label{qua731}
\end{equation}
where $I=1,\dots,n_V$ counts the number of vector multiplets and F
denotes the $N=2$ prepotential \cite{wkll,afgnt,kok,fosti}. 
Here K is the K\"ahler potential and $M_I$, $N_I$ are the"electric" 
and "magnetic" quantum
number analogs of the $N=4$ supersymmetry in heterotic string
compactifications \cite{sen}.
When $N_I$ is equal to zero then the classical spectrum 
of "electric" states can be shown to agree \cite{ropal}
with the spectrum of momentum and winding numbers in (\ref{kavou}). 
The $N=2$ K\"ahler potential is defined in terms of
$N=2$ special geometry coordinates $(X^I, i F_I)$ as
\begin{equation}
K= -\log(X^I {\bar F}_I + {\bar X}^I F_I)=-i\log(-i\Omega^{\dagger} 
\left(\begin{array}{cc}
0&{\bf 1}\\
-{\bf 1}&0\end{array}
\right)
\Omega),\;\;F_I=\frac{\partial F}{\partial X^I},
\label{qua732}
\end{equation}  
where the "period" $\Omega = (X^I, i F_I)$ defines a holomorphic section.
At the moment we will concentrate our efforts to
exhibit the dependence of the mass operator, in the perturbative
vector multiplet T, U moduli, as it originates from the 
classical mass formula (\ref{kavou}).
The dependence of the "mass" formula on
its "non-perturbative" S-part will be exhibited later.
Especially, for the case where the 
calculation of the free energy is that of the moduli space of the 
manifold $\frac{SO(2,2)}{SO(2) \times SO(2)}$, in a factorizable 
2-torus $T_{2}$, the topological\cite{fklz} bosonic free energy
is exactly the same as the one,
coming from the string one loop calculation in \cite{dkl2}. 
The total contribution to the non-perturbative superpotential, coming 
from perturbative modular orbits associated with the presence of massless 
particles of elementary string states, is connected with the existence
of the following\footnote{We calculate only the ${\sum}
{\log {\cal M}}$ since the  $\sum \log {\cal M}^{\dagger}$ quantity 
will give only the complex conjugate.}
orbits\cite{fklz,lust}
\begin{equation}
 {\Delta}_0=\sum_{2 n^t m + {q^T {\cal C} q} =2 }
{\log {\cal M}}|_{reg},\;\; 
{\Delta}_1 =\sum_{2 n^t m+{q^T{\cal C}q}=0}{\log \cal M}|_{reg}. 
\label{polop}
\end{equation} 

In the previous expressions, a regularization procedure is assumed that
takes place, which renders the final expressions finite, as infinite
sums are included in their definitions. The regularization is
responsible for the subtraction\footnote{More details 
of this precedure can be found in \cite{fklz}. } 
of a moduli independent quantity
from the infinite sum e.g ${\sum}_{n\;,m \in orbit}
{\log {\cal M}}$. 
We demand that the regularization procedure for $exp[\Delta]$
has to respect both modular invariance  and 
holomorphicity.

In eqn. (\ref{polop}), $\Delta_0$ is the   
orbit relevant for the stringy Higgs effect. This orbit is associated 
with the quantity  $2 n^T m + l^T {\cal C} l = 2$ where 
$n^T m = m_1 n^1 + 3 m_2 n^2$ for the $Z_6$-II-b orbifold.
We should note, that the term "non-decomposable " will be used in a loose 
sence, in
section four, referring to the effective supergravity theories coming from
both the non-decomposable $N=1$ orbifolds and  
 the freely acting orbifold limits, 
Enriques involutions, of $K_3$ where
the same modular subgroups of $PSL(2,Z)$, do appear \cite{kiko}.
We note that non-perturbative superpotentials invariant
under $SL(2,Z)_T \times SL(2,Z)_U$ were formulated, in terms of 
the genus two modular form ${\cal C}_{12}$, in \cite{mastiebe}.
\newline
The total contribution\footnote{We use a general embedding 
of the gauge twist in the gauge degrees of freedom.}
from the previously mentioned orbit is
\begin{eqnarray}
{\Delta}_{0}\;{\propto}\sum_{2n^T m + {l^{T}}Cl =2} \log {\cal M} 
= \sum_{n^T m =1 ,\;l^{T} C l=0} \log {\cal M} +
\sum_{n^T m =0,\;l^T C l = 2} \log {\cal M} +&\nonumber\\
\sum_{n^T m = -1,\;l^T C l = 4} \log {\cal M} +\;
\dots\;\;\;\;\;\;\;\;\;\;\;\;\;\;\;\;\;\;\;\;\;\;\;\;\;\;\;\;\;
\;\;\;\;\;\;\;\;\;\;\;\;\;\;\;\;\;\;\;\;\;\;\;\;\;\;\;\;\;\;\;\;\;
\;\;\;\;
\label{wzerooo}
\end{eqnarray}
We must notice here that we have written the sum \cite{lust} over 
the states associated with the $ SO(4,2) $ invariant orbit 
$ 2 n^T m + l^T {\cal C} l = 2 $ in terms of a sum over
$\Gamma^0(3)$ invariant orbits $n^T m = constant$ .  
The precise parametrization of the Wilson lines in the invariant
directions is not important since
it is of no interest to us in the calculation of the modular orbits.
Only the  momentum and winding number dependent part of the spin
operator
is enough for our purposes.
Remember that initially in (\ref{arxidi1}), we  
discussed the level matching condition in the case of
a $T_6$ orbifold admitting an orthogonal decomposition.
Mixing of these equations for the non-decomposable orbifolds gives us the
following equation
\begin{equation}
p_{L}^{2} - \frac{\alpha'}{2}M^{2} =  2(1-N_{L}-\frac{1}{2}
P_{L}^{2})  =  2 n^{T} m + l^{T} C l.
\label{oki1}
\end{equation}
The previous equation 
can be written in the form described in (\ref{kij1}) and 
(\ref{kij2}).
In particular it may
gives us a number of different orbits
invariant under $SO(q+2,2;Z)$ transformations. 
Namely,
$i)$ the untwisted orbit with $2 n^{T} m + l^ T C l = 2 $.
In this orbit, $ N_{L}=0,\;P_{L}^{2}=0 $. When $M^{2}=0$, this orbit is 
associated with the "stringy Higgs effect". The "stringy Higgs 
effect appears as 
a special solution of the equation (\ref{oki1}) 
at the point where $p_{L}^{2}=2$, where additional massless
particles may appear.  Or, 
$ii)$ the untwisted orbit where 
$2 n^{T} m + l^{t} C l = 0.$
Here $ 2 N_{L}+P_{L}^{2} = 2 $. This is the orbit relevant 
to the calculation of threshold corrections to the gauge couplings,
without the need of enhanced gauge symmetry points, as may happen
in the orbit $i)$. Or, 
iii) The massive untwisted orbit with 
$2N_{L} +P_{L}^{2} \geq 4$.

In this work we will always have $M^{2} \geq 0 $.
In eqn.(\ref{polop}), $\Delta_0$ is the   
orbit relevant for the stringy Higgs effect . This orbit is associated 
with
the quantity  $2 n^T m + l^T {\cal C} l = 2$ where 
$n^T m = m_1 n^1 + 3 m_2 n^2$ for the $Z_6$-II-b orbifold.

We will be first consider the contribution from the orbit
$2 n^T m + l^T {\cal C} l  = 0 $. We will be working in analogy with
calculations associated  with topological free energy 
considerations \cite{oova}.
From the second equation in (\ref{polop}), considering in general 
the $SO(4, 2)$ coset, we get for example that
\begin{eqnarray}
\Delta_{1} \propto  \sum_{n^T m + l^T C l = 0} \log {\cal M} =
\sum_{n^T m =0 ,\;l^T Cl= 0 }\log {\cal M} + 
\sum_{n^T m =-1, l^T Cl = 1}\log {\cal M} + \dots
\label{wzerooooa}
\end{eqnarray}

Consider in the beginning the term
$\sum_{n^T m =0 , l^T Cl = 0 } {\log {\cal M}} $. 
We are summing up initially the 
orbit with  $n^T m = 0; n, m \neq 0$, i.e $\Delta_1$ 
\begin{equation}
{\cal M} = 3 m_2 - im_1 U' + in^1 T + n^2 (-U' T + B C)+
{l\;dependent\;terms}. 
\label{poli}
\end{equation}
We calculate the sum over the modular orbit $n^T m + l^T C l = 0$.
As in \cite{lust} we calculate  the sum over massive
compactification states with $l^T C l\;= 0$ and $(n, m) \neq 0$. 
Namely, the orbit 
\begin{eqnarray}
\sum_{n^T m =0,\; l^T Cl = 0} \log {\cal M}=
\sum_{(n,m)\neq (0,0)}\log(3m_2 -im_1 U'+in_1 T+n_2 (-U'T))
& \nonumber\\
+\;BC \sum_{(n,m)\neq (0,0)}{\frac{n_2}{(3m_2 -im_1 U'+in_1 T-n_2 U'T)}}
+{\cal O}((BC)^2).
\label{polia}
\end{eqnarray}
The sum in relation (\ref{polia}) is topological(it excludes 
oscillator excitations) and is subject to the 
constraint $3m_2 n^2 + m_1 n^1 = 0$. 
Its solution receives contributions 
from the following sets of integers:

\begin{equation}
m_2 = r_1 r_2 \;,\;n_2 = s_1 s_2\;,\;m_1 = - 3 r_2 s_1\;,\;n_1 = r_1 s_2
\label{polib}
\end{equation} 
and
\begin{equation}
m_2 = r_1 r_2 \;,\;n_2 = s_1 s_2\;,\;m_1 = - r_2 s_1\;,\;n_1 = 3 r_1 s_2.
\label{polibb}
\end{equation}
So the sum becomes,
\begin{eqnarray}
\sum_{n^T m =0} \log {(3m_2 -im_1 U'+in_1 T-n_2 U'T)} =\sum_{(r_1,s_1) 
\neq (0,0)} \log \left(3(r_1 +is_1 U')\right)\times&\nonumber\\
\sum_{(r_2,s_2) \neq (0,0)}\log(r_2 +is_2 {\frac{T}{3}})
+\sum_{(r_1,s_1)\neq (0,0)} \log 3(r_1 +is_1{\frac{U'}{3}})
{\sum_{(r_2,s_2)\neq (0,0)}\log(r_2 +is_2 T)}&
\label{koukiol}
\end{eqnarray}

Substituting explicitly in eqn.(\ref{polia}), the values for the 
orbits in equations (\ref{polib}) 
and (\ref{polibb}) together with eqn.(\ref{koukiol}), we obtain
\begin{eqnarray}
\sum_{n^T m =0;\;l^T Cl=0} \log {\cal M}= 
\log\left({\frac{1}{3}\eta^{-2}(U^{\prime}) \eta^{-2}({\frac{T}{3}})}
\right)+
\log{\left(\frac{1}{3}\eta^{-2}({\frac{U^{\prime}}{3}})\eta^{-2}(T)
\right)}\;+&\nonumber\\
\left(BC \left( \sum_{(r_1 ,s_1 )\neq (0,0)}\frac{s_1}{r_1 +is_1 U^{\prime}}
\right) \left( \sum_{(r_2 ,s_2 )\neq (0,0)} \frac{s_2}{3(r_2 +is_2 
\frac{T}{3})}\right) \right)\;+&\nonumber\\
\left(BC \left(\sum_{(r_1 ,s_1 )\neq (0,0)}\frac{s_1}{3(r_1 +i
s_1\frac{U^{\prime}}{3})}\right)\left(\sum_{(r_2 ,s_2 )\neq (0,0)}\frac{s_2}
{r_2 +is_2 T}\right)\right)+ {\cal O}((BC)^2).&
\label{polid}
\end{eqnarray}

Notice that we used the relation
\begin{equation}
\sum_{(r_1,s_1) \neq (0,0)} \log{3}=\log\frac{1}{3}
\label{kro}
\end{equation}
with $\sum_{(r_1,s_1) \neq (0,0)} = -1$.
We substitute  
$\sum_{(r_1,s_1) \neq (0,0)} \stackrel{def}{=} \sum^{\prime}$
and $\sum_{(r_2,s_2) \neq (0,0)} \stackrel{def}{=} \sum^{\prime \prime}$.
Remember that $\sum^{\prime} \log(t_1 + i t_2 T) = 
log\;\eta^{-2}(T)$,
with $\eta(T)= exp^{\frac{-\pi T}{12}} \Pi_{n >0}(1- exp^{-2 \pi n T})$
This means that eqn.(\ref{polid}) can be rewritten as
\begin{eqnarray}
\sum_{n^T m =0;q=0} \log{\cal M}=
\log\left(\eta^{-2}(U^{\prime}) \eta^{-2}(\frac{T}{3})( \frac{1}{3})\right)+
\log{\left(\frac{1}{3}\eta^{-2}({\frac{U^{\prime}}{3}})\eta^{-2}(T)
\right)}\;+&\nonumber\\
-BC\left(\left(\partial_{U^{\prime}}\sum^{\prime}\log(r_1 + is_1 U^{\prime})
\right)\left(
\partial_{T}\sum^{\prime \prime}\log(r_2 +is_2 \frac{T}{3})\right)\right)
\nonumber\\
-BC\left(\left(\partial_{U^{\prime}}\sum^{\prime} \log(r_1 + is_1 
\frac{U^{\prime}}{3})\right)
\left(\partial_{T}\sum^{\prime \prime}\log(r_2 + is_2 T)\right)\right) + 
{\cal O}((BC)^{2}).
\label{eikos1}
\end{eqnarray}

Finally
\begin{eqnarray}
\sum_{n^T m=0;\; q = 0} \log {\cal M} = \log \left({\eta^{-2}(T)}
{\eta^{-2}(\frac{U^{\prime}}{3})})({\frac{1}{3}})\right)+
\log\left({\eta^{-2}(U^{\prime})}
{\eta^{-2}(\frac{T}{3})}{\frac{1}{3}}\right)\;&-\;&\nonumber\\
- B C\;  
\left( (\partial_T \log \eta^{-2}(T) ) ( \partial_{U^{\prime}} \log 
\eta^{-2}(\frac{U^{\prime}}{3}) ) + ( \partial_T \log 
\eta^{-2}(\frac{T}{3}) )
(\partial_{U^{\prime}} \log \eta^{-2}(U^{\prime}) )\right)&+&\nonumber\\
+\; {\cal O}((BC)^2)&&\nonumber\\
\label{kolloii}
\end{eqnarray}
So
\begin{eqnarray}
\sum_{n^T m=0;\; q = 0} \log {\cal M} = \log[\left({\eta^{-2}(T)}
{\frac{1}{3}}{\eta^{-2}{(\frac{U'}{3})}}\right)(1-\;BC\;
(\partial_T \log {\eta^{2}(T)})&\times&\nonumber\\
(\partial_U' \log {\eta^{2}(\frac{U'}{3}}))]\;\; +
\log [\;(({\eta^{-2}(U')}{\frac{1}{3}}){\eta^{-2}(\frac{T}{3})})
(1-BC(\partial_T \log &\times&\nonumber\\
{\eta^{2}(\frac{T}{3})})(\partial_U' \log\eta^{2}(U')))\;]\; +
{\cal O}((BC)^2)&&\nonumber\\
\label{plio}
\end{eqnarray}
or
\begin{eqnarray}
\sum_{n^T m=0;\; q = 0} \log {\cal M} = \log[\left({\eta^{-2}(T)}
{\frac{1}{3}}{\eta^{-2}{(\frac{U'}{3})}}\right)(1- 4\;BC\;
(\partial_T \log {\eta(T)})&\times&\nonumber\\
(\partial_U' \log {\eta (\frac{U'}{3}}))]\;\; +
\log [\;{\frac{1}{3}}(({\eta^{-2}(U')}){\eta^{-2}(\frac{T}{3})})
(1- 4BC(\partial_T \log &\times&\nonumber\\
{\eta(\frac{T}{3})})(\partial_U' \log\eta(U')))\;]\; +
{\cal O}((BC)^2)&&\nonumber\\
\label{plio1}
\end{eqnarray}
The last expression   
provides us 
with the part of the (0,2)
non-perturbative \cite{fklz,lust} generated superpotential 
${\cal W}^{non-pert}$
that comes from the
direct integration of the string massive modes. 
The non-perturbative part that depends on the dilaton will
be included later either by use of gaugino
condensates or by associating it to BPS sums.
We should say that (\ref{plio1}) may be of non-perturbative origin
only after an $e^{-S}$ dependence is included. 
The derivation of the superpotential from the sum over modular
orbits
in the case that the effective string action of decomposable 
orbifolds is invariant
under the target space duality group $SL(2,Z)_T \times SL(2,U)$ 
was found in  \cite{lust} to be the same as the 
expression argued to exist in \cite{ANTO}, for the non-perturbative 
superpotential.  
The latter was
obtained \cite{ANTO} through gaugino condensation and the requirement that 
the one loop effective action 
in the linear formulation for the dilaton be invariant\footnote{
The linear formulation is naturally present in string theory 
since the dilaton
is in the same supermultiplet with the antisymmetric tensor.}
under the full $SL(2,Z)$ symmetry up to quadratic order
in the matter fields. In exact analogy, we expect our
expression in (\ref{plio}), 
to represent the non-perturbative
superpotential of the $Z_{6} -II-b$.

Let us now exhibit the derivation of the $e^{-S}$ dependence from 
BPS and gaugino condensation generated superpotentials we have just 
comment.  We will first discuss the BPS derivation.
Consider again (\ref{qua731}).
We will attempt to generate the non-perturbative S dependence
in the case of the orbifold $Z_6-II-b$ in (\ref{kavou}).
Lets us consider transforming the period vector to the weakly 
coupled basis \cite{caf,wkll,afgnt}, where all gauge couplings
become weak at the strongly coupled limit. We transform the section, or
period vector, according to the identification
\begin{equation}
(x^I, iF_I) \rightarrow (Y^I, i L_I),
\label{asdsda12}
\end{equation}
where
\begin{equation}
Y^1=iF_1,\;L_1=i X^1,\;Y^i=X^i,\;L_i =F_i,\;for\;i=0,2,3.
\label{asdsaf54}
\end{equation}
That means
\begin{equation}  
\Omega^T = (1, TU - BC, iT, iU, iS(TU - BC),  iS, -SU, -ST)
\label{asdas1}
\end{equation}
and
\begin{eqnarray}
M_0= 3 s m_2,\; M_1= -s n_2,\;M_2 =s n_1,\;M_3 = -s m_1\nonumber\\
N_0=- p n_2,\;\;N_1 = 3 p m_2 \;\;N_2 = -p m_1,\;\;\;N_3 = p n_1.
\label{jhg12}
\end{eqnarray}
In eqn. (\ref{asdas1}, \ref{jhg12})
we assumed that the tree level $N=2$ prepotential ${\cal F}^{tree}$ in the
presence of the
Wilson lines for the "non-factorized" $T^2$ torus
is given by
\begin{equation}
{\cal F}^{tree} = -S(TU - BC),
\label{treee1}
\end{equation}
where (\ref{asdas1}) has come from the identifications
$S=iX^1/X^0$, $T=-iX^2/X^0$, and $U=-i X^3/X^0$  with the 
graviphoton $X^0=1$.

Under the identifications (\ref{jhg12}) the contribution of the 
BPS orbit that includes the perturbative orbit (\ref{polib})
factorizes and (\ref{qua731}) gives
\begin{eqnarray}
\log {\cal M} = \sum_{s,p,m_i, n_i}
 \log[(s+ ip)(3m_2 -i  m_1 U + 
in_1 T - n_2( TU - BC) ) ].
\label{orbitt1}
\end{eqnarray}
Treating in the same way the other orbit in (\ref{polibb})
we get that
the exact form of the non-perturbative effective superpotential
for the $Z_{6}$ -II-b orbifold is given by
\begin{eqnarray}
\bullet\;Z_{6} -II-b\;\stackrel{SU(3) \times SO(8)}{\rightarrow}
{\cal W}^{non-pert}_{BPS} \eta^2(S) = \eta^{-2}(T)(\frac{1}{3}){\eta^{-2}}
(\frac{U^{\prime}}{3})(1 -\;B C \;(\partial_T \log \eta^{2}
(T)) \times &&\nonumber\\ 
(\partial_U \log \eta^{2}(\frac{U^{\prime}}{3}))){\tilde W}\;+\; 
[\;({\eta^{-2}(U^{\prime})}
({\eta^{-2}(\frac{T}{3})}){\frac{1}{3}})( 1 -\;B C 
\;((\partial_T \log {\eta^{2}(\frac{T}{3}))}&&\nonumber\\
\times\;(\partial_U \log \eta^{2}(\frac{U^{\prime}}{3}))){\tilde W}
\;+\;{\cal O}((BC)^2).&&\nonumber
\end{eqnarray}
\begin{equation}
\label{secondd}
\end{equation}
The superpotential in (\ref{secondd}) transforms with modular
weight $-1$ under $SL(2,Z)_S$ S-duality transformations  
\begin{equation}
S \rightarrow \frac{a S -i b}{icS +d},\;\;\;\;\;\; ad-bc=1,
\label{sadsweq}
\end{equation}
where a,b,c,d are integers.
That means that the BPS calculation "communicates" to the
low energy heterotic string action $SL(2,Z)$ S-duality something which is
known not to be true as far as perturbation theory is concerned.
The latter can be easily recalled from gauge kinetic function arguments.
However, because (\ref{secondd}) is of non-perturbative nature 
such an expression can be compatible with perturbation theory 
only if we assume that, at the weak coupling limit,  
it matches the superpotentials that are calculated from
gaugino condensation.
So lets as an example consider a class of models\footnote{such issues
will be clarified in section four}   
that have gauge group $SU(N)$ in the "hidden" sector and involve 
matter fields with M families of quarks.
If a gauge singlet A is present that gives masses to all quarks 
then the effective superpotential arising from gaugino
condensation can be written as \cite{cacamu}
\begin{equation}
W  = {\tilde h} e^{\frac{2 \pi iS}{3N-M}},   
\label{azqws1}
\end{equation}
where $\tilde h$ depends on moduli other than S. 
Promoting the BPS expression (\ref{secondd}) into the gaugino
condensation one (\ref{azqws1}) is 
then equivalent for the following condition to hold 
\begin{equation}
\frac{3}{3N-M} = -\frac{1}{12}.
\label{azqws2}
\end{equation}
 A similar condition was found
in a different context in \cite{homou} where general forms
of S-duality invariant superpotentials were matched with 
corresponding expressions coming from gaugino condensation.

Given that at the moment there is not a non-perturbative
heterotic string formulation the issue of existence
of BPS non-perturbative generated superpotentials like in
(\ref{secondd}) has to decided
when non-perturbative heterotic calculations are available.

We will now discuss an alternative form for the
non-perturbative effective superpotential 
where the appearance of the non-perturbative dilaton  
factor is of field theoretical origin.
Its derivation originates from 
the use of the effective theory of gaugino condensates.  
In this case the non-perturbative generated superpotential 
exhibits only the leading dilaton dependence as it originates
from the the perturbative gauge kinetic function f that we have
effectively calculated through eqn.'s (\ref{polia}-\ref{plio1}). 
The appearance of the non-perturbative superpotential for gaugino 
condensation comes after integrating the contribution of the 
gaugino composite field out of the effective action
and its form will be used 
in section 5 to generate contributions to B-terms. 
It reads
\begin{eqnarray}
\bullet\;Z_{6} -II-b\;\stackrel{SU(3) \times SO(8)}{\rightarrow}
{\cal W}^{gau}e^{-3S/2b} = \eta^{-2}(T)(\frac{1}{3}){\eta^{-2}}
(\frac{U^{\prime}}{3})(1 -\;B C \;(\partial_T \log \eta^{2}
(T)) &\times&\nonumber\\ 
(\partial_U \log \eta^{2}(\frac{U^{\prime}}{3}))){\tilde W}\;+\; 
[\;({\eta^{-2}(U^{\prime})}
({\eta^{-2}(\frac{T}{3})}){\frac{1}{3}})( 1 -\;B C 
\;((\partial_T \log {\eta^{2}(\frac{T}{3}))}&\times&\nonumber\\
(\partial_U \log \eta^{2}(\frac{U^{\prime}}{3}))){\tilde W}
\;+\;{\cal O}((BC)^2),&&\nonumber\
\end{eqnarray}
\begin{equation}
\label{first}
\end{equation}
where S is the dilaton and b the $\beta$-function of the
condensing gauge group, e.g for an $E_8$ hidden group $b=-90$, 
and ${\tilde W}$ depends on the moduli of 
the other planes, e.g the third invariant complex plane, when 
there is no cancellation of anomalies by the Green-Schwarz 
mechanism. 
For example for the Z$_6$-II-b orbifold, if $T_3$ is the moduli of the
third complex plane then
\begin{equation}
{\tilde W}= \frac{1}{\eta^2(T_3)}.
\label{other1} 
\end{equation}
The previous discussion was restricted to small values of the 
Wilson lines where our $(0,2)$ orbifold goes into $(2,2)$.
The grouping of terms in the form presented 
in (\ref{plio1}) is our natural choice. In this form $\cal W$
the two additive factors in eqn.(\ref{first}) have separately the
invariances of the K\"ahler potential in the linear representation
of the dilaton. In other words, candidates for
$\cal W$ for the $Z_6 -II-b$, or some other heterotic string
compactification having
the same duality symmetry group,
are either the sum of 
the individual factors in
eqn.(\ref{first}) or its factor separately.
Specifically, grouping together the first with the third term and the 
second with the forth term we get the result (\ref{plio1}). 
On the other hand,
any other regrouping of terms in (\ref{kolloii}) is excluded as a candidate
for $\cal W$ at it does not have the correct modular 
weight\footnote{Remember, that we have changed the notation from
$U^{\prime}$
to U.}. Similar results hold for the other orbifolds.

For the $Z_4 - a$ orbifold defined by the action of the complex twist
$\Theta=(i,i,-1)$ on the lattice $SU(4) \times SU(4)$, the mass operator
for the $\Theta^2$ subspace is
\begin{equation}
m^2 = \frac{1}{2T_2 U_2} |T U n^2 + T n^1 -2 U m_1 +2 m_2|^2.
\label{sxolio2}
\end{equation}
and it is invariant under the $\Gamma_o(2)_T \times SL(2,Z)_U$ target 
space duality modular group.
The spin is
\begin{equation}
n^T m = 2 m_1 n^1 + 2 m_2 n^2.
\label{sxolio2}
\end{equation}
Considering, as before, a general embedding of the Wilson lines in the 
gauge degrees of freedom we get
\begin{eqnarray}
\sum_{n^T m =0,\; q = 0} \log {\cal M}=
\sum_{(n,m)\neq (0,0)}\log(2m_2 -2im_1 U'+in_1 T+n_2 (-U'T))
& \nonumber\\
+\;BC \sum_{(n,m)\neq (0,0)}{\frac{n_2}{(2m_2 - 2im_1 U'+in_1 T-n_2 U'T)}}
+{\cal O}((BC)^2).
\label{sxolio3}
\end{eqnarray}
The topological sum constraint $2m_2 n^2 + 2m_1 n^1 = 0$ receives 
contributions from the following sets of integers:
\begin{equation}
m_2 = r_1 r_2 \;,\;n_2 = s_1 s_2\;,\;m_1 = -  r_2 s_1\;,\;n_1 = r_1 s_2.
\label{polibbbb}
\end{equation} 
Expanding, 
\begin{eqnarray}
\log {\cal M} = \sum_{n^T m=0;q=0} \log((r_1 + is_1 U)2(r_2 + i 
\frac{T}{2} s_2))\; + &\nonumber\\
BC \left( \sum^{\prime} \frac{s_1}{r_1 + is_1 U}
\right) \left( \sum^{\prime \prime} \frac{s_2}{2 (s_2 + i 
\frac{T_2}{2} s_2)} \right) + {\cal O}((BC)^2).
\label{polibvc}
\end{eqnarray}
Explicitly,
\begin{equation}
\log {\cal M}= \log(\eta^{-2}(U) \frac{1}{2} \eta^{-2}(\frac{T}{2})) 
-BC\sum^{\prime} (\partial_U \log(r_1 + is_1 U)) \sum^{\prime \prime}
(\partial_T \log((r_2 + i \frac{T}{2} s_2))) + {\cal O}((BC)^2),
\label{polibvcc}
\end{equation}
and
\begin{equation}
\log {\cal M}=\log(\eta^{-2}(U) \frac{1}{2} \eta^{-2}(\frac{T}{2}))
-BC \left( (\partial_U \log \eta^{-2}(U))( 
\partial_T \log  \eta^{-2} (\frac{T}{2})) \right)
+ {\cal O}((BC)^2).
\label{polibvcccc}
\end{equation}
Finally, ${\cal W}^{pert}$ becomes 
\begin{eqnarray}
\bullet\;Z_4 -a\;\stackrel{SU(4)\times SU(4)}{\rightarrow}\;{\cal W}^{pert}
=(\eta^{-2}(U)\frac{1}{2}\eta^{-2}(\frac{T}{2}))
(1 -4BC(\partial_U \log \eta(U))
(\partial_T \log\eta(\frac{T}{2}))+ &\nonumber\\
{\cal O}((BC)^2).
\label{polibvcccca}
\end{eqnarray}
For the $Z_8 -II-a$ orbifold defined by the action of the complex twist 
$\Theta = exp[2 \pi i(1,3,-4)/8]$ on the torus lattice $SU(2) \times 
SO(10)$,
the mass operator for the $\Theta^2$ subspace is given by
\begin{equation}
m^2 = \frac{1}{2 T_2 U_2} |(TU-BC) n^2 + Tn_1 -2 U m_1 + m_2 + l\;dep.\;
terms|^2
\label{sdcaq}
\end{equation}
and it is invariant under the modular group $\Gamma_o(2)_T \times 
\Gamma^o(2)_U$  when $B=C=l=0$.
The superpotential $\cal W$, receives contributions from
the orbit
\begin{equation}
n^T m = 2 m_1 n_1 + m_2 n_2,\;l=0 
\label{sdcaqq1}
\end{equation}
The latter can be solved by decomposing it in the two
inequivalent orbits
\begin{eqnarray}
m_2 = 2 r_1 r_2,\; n_2 =s_1 s_2,\;m_1 =-r_2 s_1,\;n_1 = r_1 s_2&\nonumber\\
m_2 =  r_1 r_2,\; n_2 =2s_1 s_2,\;m_1 =r_2 s_1,\;n_1 = -r_1 s_2.
\label{sdcaqq11}
\end{eqnarray}
Summing over the orbits we get that $\cal W$ is
\begin{eqnarray}
\bullet\;Z_8 -II-a\;\stackrel{SU(2) \times SO(10)}{\rightarrow}\;\;
{\cal W}= \left( \frac{1}{2} \eta^{-2}(\frac{T}{2})\eta^{-2}(U)
\right) (1 - 4 BC(\partial_{T}\log\eta(\frac{T}{2}))&\nonumber\\
\times
(\partial_{U}\log\eta (U)))
+ \left( \eta^{-2}(T) \eta^{-2}(2U)\right) \left( 1 - 4 BC(\partial_{T}
\log \eta^{-2}(T) ) (\partial_{U} \log \eta^{-2}(2U) )
\right),\;\;\;\;\;&\nonumber\\
\label{sdcaqq111}
\end{eqnarray}
while the non-perturbative generated superpotential coming
from gaugino condensation is given by
\begin{equation}
{\cal W}^{non-pert} = e^{3S/2b} {\cal W}|_{Z_8 -II -a}
\label{lilio}
\end{equation}
The same form for $\cal W$ as in (\ref{sdcaqq111}) can be shown to hold 
for the non-decomposable
$Z_4 -b$ orbifold defined by the action of the complex twist
$\Theta = (1,1,-4)/2$
on the six dimensional torus lattice $SU(4) \times SO(5) \times SU(2)$.
The latter has the same modular symmetry group as the $Z_8 -II-a$ when 
$B=C=0$.

For the orbifold, listed as $Z_6 -II-c$, and defined by the
action of the complex twist $\Theta=[(2,1,-3)/6]$, on the lattice
$SU(3) \times SO(7) \times SU(2)$, we derive $\cal W$ as
\begin{eqnarray}
\bullet\;Z_6 -II-c\;\stackrel{SU(3)\times SO(7)\times
SU(2)}{\rightarrow}
\;{\cal W}= \left(\frac{1}{3}\eta^{-2}(\frac{T}{3})\eta^{-2}(U)
\right) [1 - 4 BC (\partial_{T} \log \eta(\frac{T}{3}))
\;\times\;&\nonumber\\
 ( \partial_U \log\eta (U) ) ] +
\left(\eta^{-2}(T)\eta^{-2}(3U)\right)\left(1 - 4BC(\partial_{T}
\log \eta(T))(\partial_{U} \log \eta (3U))\right).\;\;\;\;\;\;\;\;\;\;\;&
\label{sdcaqq1111}
\end{eqnarray}
The modular group for the latter orbifold is 
$\Gamma^o(3)_T \times \Gamma_o(3)_U$ when $B=C=0$.

\section{Non-perturbative Superpotentials from Generalized Coxeter 
Orbifolds} 

We will end the discusion of non-perturbative superpotentials
by calculating $\cal W$ for generalized Coxeter orbifolds (GCO).
We will present the discussion for the CGO $Z_8$ orbifolds, 
defined\footnote{
This orbifold was included in the classification list of \cite{erkl}.
However, neither its moduli dependent gauge couplings nor 
its topological free energy were studied in \cite{deko}.
Here, we calculate its free energy. The moduli dependence of the gauge 
couplings were treated in \cite{kokos0}.}
by the Coxeter twist
$(e^{\frac{i \pi}{4}}, e^{\frac{3 i \pi}{4}}, -1)$ on the root lattice
of $A_3 \times A_3$. We remind that this orbifold is non-decomposable in
the sence that the action on the lattice twist on the $T_6$ torus does not 
decompose into the orthogonal sum $T_6 = T_2 \oplus T_4$ with the fixed plane 
lying on $T_2$ torus.  For this orbifold, the twist can be equivalently be 
realized through the generalized Coxeter automorphism $S_1 S_2 S_3 P_{35}
P_{36}P_{45}$. In general, the GCO is defined as a product of the Weyl 
reflections $S_i$ of the simple roots and the outer automorphisms 
represented by the transposition of the roots. An outer automorphism 
represented by a transposition which exchange the roots 
$ i \leftrightarrow j$, is denoted by $P_{ij}$ and is a symmetry of the 
Dynkin diagram. 
The realization of the point group is generated by
\begin{eqnarray}
Q=\left(\begin{array}{cccccr}
0 & 0 & 0 & 0 & 0& -1\\
1 & 0 & 0 & 0 & 0&  0\\
0 & 1 & 0 & 0 & 0& -1\\
0 & 0 & 1 & 0 & 0& 0\\
0 & 0 & 0 & 1 & 0& -1\\
0 & 0 & 0 & 0 & 1& 0
\end{array}\right).
\label{aeqn1}
\end{eqnarray}
Threfore the metric defined by $g_{ij} = <e_i|e_j>$  has three and the 
antisymmetric tensor b an other three deformations. The
threshold corrections, topological free energy and effectively the 
$\cal  W$ will depend on the moduli of the unrotated complex plane.
That is if the action of the generator of the point group leaves some 
complex plane unrotated the equations 
\begin{equation}
gQ = Qg,\;\;bQ =Qb
\label{asdwq1}
\end{equation}
determine the background fields in terms of the independent
deformation parameters. 
Solving these equations one obtains for the metric
\begin{eqnarray}
G =\left(\begin{array}{cccccr}
R^2& u& v& -u&-2u-R^2 &-u\\
u & R^2 & u &v &-u & -2v-R^2\\
v & u & R^2 &u &v &-u\\
-u& v & u &R^2 &u &v\\
-2v-R^2&-u &v &u &R^2 &u \\
-u& -2v-R^2&-u& v& u& R^2
\end{array}\right),
\label{asdqw12}
\end{eqnarray}
with R, u, v $\in {\cal R}$ and the antisymmetric tensor field :
\begin{eqnarray}
B= \left(\begin{array}{cccccr}
0&x&z&y&0&-y\\
-x&0&x&z&y&0\\
-z&-x&0&x&z&y\\
-y&-z&-x&0&x&z\\
0&-y&-z&-x&0&x\\
y&0-y&-z&-x&0
\end{array}\right).
\label{asdqw123}
\end{eqnarray}
The $N=2$ orbit for these sectors will contain completely unrotated planes, 
${\cal O}= (1, \Theta^4), (\theta^4, 1),
\newline
 (\theta^4, \theta^4)$.
Consider now the usual parametrization of the $T^2$ torus
with the $(1,1)$ T-modulus and the $(2,1)$ U-modulus as,
\begin{eqnarray}
T = T_1 + iT_2 = 2(b + i \sqrt{det g_{\perp}} ) & \nonumber\\
U = U_1 + i U_2 = \frac{1}{g_{\perp 11}} (g_{\perp 12} + i \sqrt{det g_{\perp}}).
\label{asdawq23}
\end{eqnarray}
Here, $g_{\perp}$ is quiquely determined by 
$w^t Gw = (n^1 n^2) g_{\perp} \left(\begin{array}{c}n^1\\n^2\end{array}
\right)$.
Here, b is the value of the $B_{12}$ element of the two-dimensional matrix
of the antisymmetric field B. This way one gets
\begin{equation}
T= 4(x-y) + i 8v,
\label{zeta1}
\end{equation}
\begin{equation}
U = i.
\label{zeta2}
\end{equation}
The mass operator in the $(1,\Theta^4)$ untwisted subspace takes the form 
\begin{equation}
m^2 = \frac{1}{2 T_2 U_2} |-TUn_2 + iTn^1-2Um_1 + 2 m_2|^2
\label{poli1}
\end{equation}
but with the values of T, U given by eqn's (\ref{zeta1}, \ref{zeta2}).
Because, $n^T m = 2 (m_1 n^1 + m_2 n^2) + l-terms $, $\cal W$ takes
exactly the same form as in eqn. (\ref{polibvcccca}) implemented by the
exponential dilaton $\propto e^{-3S/2b}$ factor.

Here we can make a comment related to the contribution from the first 
equation
in (\ref{polop}) which is relevant to the stringy Higgs effect.  
Take for example the expansion (\ref{wzerooo}). Let's examine the first 
orbit
corresponding to the sum ${\Delta}_{0,0} =\sum_{n^T m =1, q=0} 
\log {\cal M}$ . This orbit is the orbit for which some of the 
previously 
massive states, now become massless. At these points
${\Delta}_{0,0}$
has to exhibit the logarithmic singularity. 
In principle we could predict, in the simplest case when the Wilson 
lines have been switched off the form of ${\Delta}_{0,0}$.
The exact form, when it will be calculated has to respect that that
the 
quantity $e^{\Delta_{0,0}} $ has modular\footnote{This point
was not explained in \cite{lust} but it is obvious that it 
corresponds to the superpotential and thus transforming with
modular weight $-1$.}
weight $-1$ and reflects exactly the 
presence of the physical singularities of the theory.
We will not attempt to calculate this orbit as its sum 
is largely unknown.

\section{Solutions to the $\mu$ Problem}

In the previous section, we have calculated contributions to the 
non-perturbative superpotentials that are coming from the 
sum over modular orbits of massive untwisted states present
in non-decomposable orbifold compactifications. The latter 
vacua represent N=1 $(0, 1)$ heterotic compactifications
based on the orbifolds $T^6/Z_{\nu}$.
The non-perturbative superpotentials take the form of eqn. (\ref{dgggin}).
The fields B, C represent matter fields. If we identify the matter fields
B, C associated with the untwisted complex plane,
 with the Higgs fields of the minimal supersymmetric standard model 
the $\mu$-terms are exactly given by eqn. (\ref{term}) without any 
contributions from higher weight interactions.  

The contributions from the first two terms
in (\ref{term}) are the standard contributions present in effective  
supergravity theories irrespectively of the origin of supergravity theory. 
The tree level expression for the function H in eqn. (\ref{term}) is given
by
\begin{equation}
H_{BC} = \frac{1}{(T + {\bar T})(U + {\bar U})}.
\label{treebar1}
\end{equation}
Later we will provide the $\tilde \mu$ contributions of the third term in
(\ref{term}) as its contribution can be read from the 
non-perturbative superpotential.
The $\tilde \mu$ contributions to the 
$\mu$-term can be determined once ${\cal W}_{BC}$ is singled out.
Let us discuss how the form of the non-perturbative superpotentials 
calculated on the previous section solves the $\mu$ problem.
The central element of our "proof" will be to justify to presence of an
effective 
$\mu_{BC} BC$ non-zero mass term after supersymmetry breaking, at low 
energies, in the 
observable sector associated with the presence of the $\tilde \mu$-term
in (\ref{term}). Lets us for the moment suppose that 
$\tilde \mu =0 $.      
Let me first take the contributions of the first two terms in eqn.
(\ref{term}). Their presence has been discussed in \cite{gima,casa}. 
Their origin resides on the presence of the mixing $H_{BC}$ term in 
the K\"ahler potential as in (\ref{papa}) e.g
\begin{equation}
\hat{K} = \dots +  H_{BC} B C ;\;\;W= W_o.
\label{kiolo}
\end{equation}
The effect of this term in the effective theory can be shown to be 
equivalent to the effect that is coming from the 
K\"ahler potential $K^{\prime}$ and 
superpotential $W^{\prime}$, in the lowest order of expansion in the 
matter fields, given by
\begin{eqnarray}  
K^{\prime}= \hat{K}- H_{BC} BC,\; W^{\prime} = W_o e^{H_{BC} BC} 
\approx W_o(1 + H_{BC} BC)|_{lowest \;order},& &\\
G=K^{\prime} + \ln |W^{\prime}|^2.\;\;\;\;\;\;\;\;\;\;\;\;\;\;\;\;\;
\;\;\;\;\;\;\;\;\;\;\; &&
\label{werds1}
\end{eqnarray}
That means that the effective theories coming from (\ref{werds1}) 
and (\ref{kiolo}) have the same G-function.
Especially to the lowest order of expansion in the matter fields
the theory with the Higgs mixing term in the K\"ahler potential, eqn.
(\ref{kiolo}) 
becomes equivalent to the one coming from the expansion of the 
superpotential in (\ref{werds1}).  
If we assume vanishing of the cosmological constant, (\ref{werds1})  
gives rise to the generation of 
mass terms in the low energy potential that look
like as\footnote{see for example \cite{casa}.}
\begin{eqnarray}
V \propto m_{3/2}^2 (1 + \kappa)^2 |B|^2 +  
m_{3/2}^2(1 + \kappa)^2 |C|^2 + 2m_3^2 \kappa (BC + h.c)
\propto -\hat{\mu}^2 (BC + h.c),\\
m_{3/2}^2=e^{K}W_o,\,\;\;\;\;\hat{\mu} = 2m_{3/2}e^{K/2}\mu . 
\;\;\;\;\;\;\;\;\;\;\;\;\;\;\;\;\;\;\;\;\;\;\;\;\;\;
\label{poten}
\end{eqnarray}
The previous potential gives the usual Higgs potential of the 
supersymmetric standard model when it rewritten in the form
\begin{equation}
V = \dots \mu_1^2 |B|^2 +\mu_2^2 |C|^2 + B_{\mu} m_{3/2} e^{\frac{1}{2}K}
\mu.
\label{superpo}
\end{equation}
That means 
that the effective $\mu$-term is created in the observable superpotential, 
that includes
MSSM model couplings, from the following expression  
\begin{equation}
\mu BC,\;\; \mu =  <W_o> H_{BC} \equiv W_{BC}.
\label{effeccti}
\end{equation}    
We conclude that the presence of the mixing term $W_{BC}$ in the 
superpotential of the theory has the same effect in the low energy
spectrum of the observable theory as the one coming from the 
presence of the $H_{BC}$ term in the K\"ahler potential. They both 
generate dynamically the same $\mu$-term presence in the effective theory.

Take now for example $\cal W$ found for the $Z_4$ - a orbifold in eqn.
(\ref{polibvcccca}). Now
${\cal W}$ takes the form 
\begin{equation}
W|_{Z_4 \times Z_4}  = W_o(T, U) + \kappa(T, U) W_o(T, U) B C 
+ {\cal O}((BC)^2)
\label{reds1}
\end{equation}
with
\begin{equation} 
W_o (T, U) = \frac{1}{2}\eta^{-2}(U) \eta^{-2}(\frac{T}{2}),\;\;\;\;\;
\;\kappa(T, U) = - 4 (
\partial_U \log \eta(U))(\partial_T \log \eta(\frac{T}{2})).
\label{solutio1}
\end{equation}

By comparison of (\ref{polibvcccca}), (\ref{poten}) and ({\ref{solutio1}) 
the following identifications follow
\begin{equation}
 W = W_o + \kappa W_o BC,\;\;\;\;\mu \equiv\kappa <W _o> ,
\;\;\;W= W_o + \mu BC.
\label{epitelo1}
\end{equation}
In other words, the form of the string theory superpotential in 
(\ref{epitelo1}) generates naturally the form of the effective
superpotential that is needed in order to produce the mixing effect of 
an effective $\mu$-term in the low energy superpotential of the theory.  
As we have already said the form of the string theory generated
superpotential is consistent with the invariance of the one loop corrected
effective action, in the linear representation of the dilaton, under 
the target space duality transformations
that leave the tree level K\"ahler potential (\ref{ertklio}) invariant.

We should always remember at this point that  
the condensation superpotential may
be used to break supersymmetry at a scale smaller than $M_{Plank}$ and to
generate masses to 
quark fields in the presence of the Higgs fields B, C.
Lets us now justify the connection of the non-perturbative 
superpotentials calculated previously to solutions of the 
$\mu$-problem.
Lets us consider for this reason that the effective theory of light modes
is coming from
a superstring vacuum which has a YM gauge group $SU(N)$ and 
in addition M $SU(N)$ "quarkslike" 
$Q_a$ , $a=1, \dots,M$ fields 
that are given mass by the same $SU(N)$ singlet field A. 
The mass term in this case reads
\begin{equation}
W^{pert} = -\sum_{a=1}^M A Q_a {\bar Q}_a .
\label{opli12}
\end{equation}
After applying a standard procedure \cite{lutay} the
non-perturbative superpotential from gaugino condensates (NPS)
appears in the form
\begin{equation}
W^{non-pert} \propto {(det \Pi)^{1/N}}e^{3S/2b},
\label{opli14}
\end{equation}
with $\Pi$ a mass matrix defined as
\begin{equation}
det \Pi= A^{M}.
\label{opli15}
\end{equation}
\newline
Assume now that the coupling $A B C Q_a {\bar Q}_a$ is allowed 
then the superpotential (\ref{opli12}) may be promoted to
to the
form \cite{veya,lutay}
\begin{equation}
W^{pert} =  -\sum_{a=1}^{M}A (1+ \kappa^{\prime}BC) Q_a {\bar Q}_a ,
\label{loioasda}
\end{equation}
Comparing (\ref{opli14}, \ref{opli15}, \ref{loioasda})
we derive
\begin{equation}
W^{non-pert}  \propto {(det \Pi^{\prime})}^{1/N} e^{3S/2b},
\label{loios1}
\end{equation}
where the mass matrix ${\Pi}^{\prime}$ is given by
\begin{equation}
det \Pi^{\prime} = A^M ( 1+ \kappa^{\prime} BC)^M.
\label{loios2}
\end{equation}

The usefulness of the non-perturbative superpotential found for the 
$Z_4 -b$ $N=1$ orbifold   
relies on the fact that when for a 
particular vacuum the coupling $A B C Q_a {\bar Q}_a$ is allowed, 
signalling the presence of a non-renormalizable term,
then we 
can identify 
\begin{equation}
W^{non-pert} \approx A^{\frac{M}{N}}(1 + \kappa^{\prime} \frac{M}{N}
BC)e ^{3 S/2b} \equiv W^{non-pert}|_{Z_4 -b} .    
\label{tion1}
\end{equation}
Comparison between (\ref{tion1}) and (\ref{solutio1}) for the $Z_4$-b
suggests the identification 
\begin{equation}
A^{\frac{M}{N}} = W_o(T,U), \;\;\kappa^{\prime} \frac{M}{N}=\kappa(T, U),  
\label{tion2}
\end{equation}
so that the "hidden" field A is a function of the moduli fields T, U.

We also note that the results (\ref{tion1}, \ref{tion2})  
presuppose in their derivation the fact that
at the limit $S \rightarrow \infty $ and $M_{Plank} \rightarrow \infty$ the
NPS superpotentials
fix the dilaton at its acceptable weak coupling value and are given by a
truncation procedure that identifies them with the superpotentials
appearing in the globally supersymmetric QCD in the form (\ref{opli14}).

For the orbifold $Z_6 -II-b$ from (\ref{first})  
the $\bar \mu$ contributions 
to the $\mu$ term read
\begin{eqnarray}
\bullet Z_6 -II-b\;\stackrel{SU(3) \times SO(8)}{\rightarrow} 
{\tilde \mu}e^{-G/2}e^{-3S/2b}=
[\;( -\eta^{-2}(T) \frac{1}{3} 
(\eta^{-2} \frac{U^{\prime}}{3}))(\partial_T \log {\eta^{2}
(T))}(\partial_U \log &\times&\nonumber\\
\eta^{2}(\frac{U}{3}))\;]{\tilde W}\;+\;- [\;({\eta^{-2}(U)}
({\eta^{-2}(\frac{T}{3})}){\frac{1}{3}})(
\;((\partial_T \log {\eta^{2}(T))}&\times&\nonumber\\
(\partial_U \log \eta^{2}(\frac{U}{3})))\;]{\tilde W}
\;+\;{\cal O}((BC)^2).\;\;\;\;\;\;\;\;\;\;\;\;&&\nonumber\\
\label{lopolo}
\end{eqnarray}
As we have said each of the factors in eqn. (\ref{lopolo}) can be used
as a possible $\tilde \mu$-term contribution.
The latter forms exclude the ansatz for the $\tilde \mu$-term used 
in \cite{Love2} in the context of CP violation.

The expressions of the
$\tilde \mu$ terms for the classes of non-decomposable orbifolds 
of the orbifolds appearing in sections 2 and 3 may be summarized as
\newline
$\bullet Z_8 |_{A_3 \times A_3}{\rightarrow}
{\tilde \mu}e^{-3S/2b} e^{- G/2}=-\frac{1}{2}\eta^{-2}(\frac{T}{2})\eta^{-2}(U)
\left( \partial_T\log(\eta^2(\frac{T}{2}))\right)
\left(\partial_U \log\eta^{2}(U)\right),$
\newline
\begin{equation}
\label{tlikosi2}
\end{equation}
$\bullet  Z_4 - a |_{SU(4) \times SU(4)}$ the same value of 
${\tilde \mu}$-term as the $Z_8|_{A_3 \times A_3}$
\begin{eqnarray}
\bullet\;Z_4 -b|_{SU(4) \times
SO(5) \times SU(2)}
{\rightarrow}
{\tilde \mu}e^{-3S/2b}e^{- G/2}=- [\frac{1}{2} \eta^{-2}(\frac{T}{2})
\times \eta^{-2}(U) ] (\partial_{T}\log\eta^2(\frac{T}{2})\;\;\;\;\;\;
\;\;\;&\nonumber\\
\times\;(\partial_{U}\log\eta^2 (U))
- \eta^{-2}(T)\eta^{-2}(2U)(\partial_{T}\log \eta^{2}(T))(\partial_{U}
\log\eta^{2}(2U)),&\nonumber\\
\label{sdcaqq1112}
\end{eqnarray}
\newline
$\bullet Z_8 -II-a |_{SU(2) \times SO(10)};\;$ the same value of 
$\tilde \mu$-term as the ${Z_4-b}|_{SU(4) \times SO(5) \times SU(2)}$
\begin{eqnarray}
\bullet\;Z_6 -II-c|_{SU(3) \times SO(7) \times SU(2)}{\rightarrow}
\;{\tilde \mu}e^{-3S/2b}e^{- G/2}=-\left(\frac{1}{3}\eta^{-2}
(\frac{T}{3})\eta^{-2}(U)\right)(\partial_{T} \log \eta^2(\frac{T}{3}))
\;\times\;&\nonumber\\
(\partial_{U}\log\eta^2(U))-
\eta^{-2}(T)\eta^{-2}(3U)(\partial_{T}
\log \eta^{2}(T))(\partial_{U} \log \eta^2(3U)).&\nonumber
\end{eqnarray}
\begin{equation}
\label{sdcaqq11123}
\end{equation}

\section{The $\mu$-terms in M-theory}

$\bullet$ {\em Different scenarios}
\newline
The purpose of this section is to 
examine some phenomenological consequences of our approach
and in particular
to show in which way it is 
possible to convert our $\mu$-terms solutions of the previous section
to their M-theory equivalents.
In the context of M-theory
Horava and Witten\cite{howi} proposed that the strong coupling limit of 
the $E_8
\times E_8$ heterotic string is described by eleven dimensional 
supergravity on a manifold with a boundary.  The two $E_8$'s,
"considered" as the observable and hidden sectors of the 
early supergravity theories, live on the opposite ends of the 
squased $K_3$ surface, the line segment $S^1/Z_2$. 
Further compactification of M-theory to four 
dimensions revealed a 
number of interestring features including computation of K\"ahler 
potential \cite{owa}, and supersymmetry breaking terms 
\cite{ducro,anqu}.
The low energy supergravity which describes 
the theory
in four dimensions has been discussed in several works 
\cite{nano,li,latho,anqu,ducro}, 
while
soft supersymmetry breaking terms like gaugino masses $M_{1/2}$,
scalar masses $m_0$, and A-terms have been calculated in
\cite{chkimi,nils}. We are particularly interested in the
B-soft supersymmetry breaking term. The latter 
depends on the details of the mechanism that generates the
$\mu$-terms. Because of the non-perturbative nature of our solutions
we will further assume, initially, that the B-soft term is the result of  
non-perturbative $\mu$-term generation.

$\bullet$ {\em M-theory B-terms flowing to $N=1$ orbifolds}

We are interested to calculate the value of the M-theory B-term that 
its weakly coupled limit flows to the limit of B-terms associated with
the $N=1$ orbifolds \cite{sce}.
In this case the M-theory B-term reads\footnote{
Note that a different form of the B-soft term was appeared in
\cite{bakralo}.
We disagree with the form of the 
B-term presented in this work.} 
\begin{eqnarray}
B_{\mu}= \left(-3{\tilde C} cos\theta e^{-ia_S} -\sqrt{3}{\tilde C} 
sin\theta
+\frac{6{\tilde C} cos\theta (S + {\bar S})}{3(S+{\bar S})+ \alpha 
(T + {\bar T})}
+ \frac{2\sqrt{3}{\tilde C} sin\theta \alpha (T + {\bar T})}
{3 (S + {\bar S}) +
\alpha (T + {\bar T})} -1 \right) &\nonumber\\
+\;\;F^S \partial_S \ln \mu^{M} + F^T \partial_T \ln
\mu^{M},&\nonumber  
\end{eqnarray}
\begin{equation}
\label{akliopa}
\end{equation}
where\footnote{we see in the following that this form
of B-term is compatible with the M-theory limit of certain 
decomposable orbifold compactifications of the heterotic string}   
\begin{equation}
F^S= \sqrt{3} m_{3/2} {\tilde C} (S +{\bar S})sin\theta,\;\;
F^T=m_{3/2} {\tilde C}(T +{\bar T})cos\theta,\;\;{\tilde C}=1 + 
\frac{V_o}{3m^3_{3/2}},   
\label{akliopa1}
\end{equation}
$V_o$ the value of the cosmological constant and $F^S$, $F^T$
 the dilaton and moduli auxiliary fields. 
In addition, we have used the K\"ahler function K of  M-theory
\begin{eqnarray}
K= -\ln(S + {\bar S}) -3 \ln(T + {\bar T})+
\left( \frac{3}{T + {\bar T}} + \frac{\alpha}{S + {\bar S}}\right)
|C|^2\\
f_{E_6}= S + \alpha T,\;\;f_{E_8} = S - \alpha T,\;\;W=d_{pqr}
C^p C^q C^r,
\label{koilopi}
\end{eqnarray}
the gauge kinetic functions f for the observable and hidden sector (2,2)
model, and finally the perturbative superpotential W. The latter is 
a function of the matter fields C.
In the derivation of eqn. (\ref{akliopa}) we have used the relation
\begin{eqnarray}
B_{\mu}^{M-theory} =
m_{3/2}[-1 +{\tilde C} \sqrt{3} sin \theta (K_{o{\bar S}}^S)^{-1/2}
\left(K_o^S + \frac{\mu_S}{\mu} - \frac{{\tilde K}_C^S}{{\tilde K}_C}
- \frac{{\tilde K}_{\bar C}^S}{{\tilde K}_{\bar C}}
\right) +\nonumber\\ 
{\tilde C} \sqrt{3} cos\theta (K_{o{\bar T}}^T)^{-1/2}
\left(K_o^T + \frac{\mu_T}{\mu} -\frac{{\tilde K}_C^T}{{\tilde K}_C}
- \frac{{\tilde K}_{\bar C}^T}{{\tilde K}_{\bar C}}\right)],
\label{correctb}
\end{eqnarray}
which is a generalization of the standard relations
existing for the B-terms of $N=1$ four dimensional orbifolds \cite{bim}.
In this way, even if the relation between B-term and $\mu$-term is fixed in 
supergravity it is made compatible to include changes in the moduli
metrics.
Note that by $K_o$ we denote that part of the K\"ahler potential 
containing the moduli metrics while $ \tilde K$ is the part involving
matter fields. 
\newline
The B-soft term involving the $ F^i\partial_i \ln \mu$ term 
has modular weight $-1$ as it should.
In this form it can be easily proved that the M-theory B-soft term
eqn.(\ref{akliopa})
flows into its weakly coupled heterotic limit,
e.g by taking appropriate limits in eqn.(2.19) of \cite{bim}, 
when $\alpha (T + {\bar
T}) <<(S + {\bar S})$. At the latter limit the B and the rest of the
soft terms flow to their large Calabi-Yau limit equivalent to the
blow up of twisted moduli fields of abelian (2,2) orbifold
compactifications counterparts \cite{bim}.    
The form of the M-theory $\mu$-term in (\ref{akliopa}) may be 
replaced by the
heterotic $\mu$-terms, of sect. 4, only when 
\begin{equation}
Re(S) >>[4 \pi^2Re(T)]^3,\;\;Re(T) >>\frac{1}{4 \pi^2}
\label{akliopa2}
\end{equation}
When (\ref{akliopa2}) holds we are in the heterotic limit so that
starting with our heterotic string $\mu$-term solution of sect. 4
and by varying Re(S) while keeping Re(T) fixed we can extrapolate
smoothly from heterotic string to M-theory \cite{chkimi}.
We should note that retaining the weak perturbative properties
of our theory demands in addition that $(S+{\bar S})+ \alpha(T 
+{\bar T}) \approx 4$ or in other words 
$0 < \alpha(T + {\bar T}) \leq 2$.

We shall make a comment at this point regarding the validity
of the effective theory coming from the specific compactification
of the M-theory \cite{owa} that gives us the quantities in (\ref{koilopi})  
and its compatibility with the existing $N=1$ orbifold compactifications
of the $N=1$ heterotic string \cite{dhvw}. Remember that the
three $T_i$, $i=1,2,3$ moduli are all considered in the same footing
in (\ref{koilopi}) i.e $T_1=T_2=T_3=T$. 
Because the definition of the $\mu$-term in (\ref{akliopa}) involves at 
least one 
U moduli it is valid for those B-terms that flow, at  
their weakly coupled limit,  to the $N=1$ 
orbifolds having at least one U-moduli in the e.g 
third complex plane. The latter
orbifolds include the $Z_N$ one's based
 on $Z_4$, $Z_6$, $Z_8$, $Z_{12}^{\prime}$ and the $Z_N 
\times Z_M$ one's based on $Z_2 \times Z_4$ and $Z_2 \times Z_6$.

Take for example the $Z_2 \times Z_6$ orbifold. 
This orbifold has three $N=2$ $T_1$, $T_2$, $T_3$ and one $U_3$ moduli.
The 
non-perturbative superpotential associated with the $T_3$, $U_3$ moduli
is given by
\begin{equation}
W_{Z_2 \times Z_6}^{non-pert}= e^{-3S/2b} \eta^{-2}(U_3) \eta^{-2}(T_3)
\left(1 - AB (\partial_{T_3} \ln(\eta^2 (T_3)))(\partial_{U_3} \ln(\eta^2
(U_3)))
\right).
\label{asdfgh}
\end{equation}
Lifting this $\mu$-term, where both T, U moduli are involved, to the one 
coming from M-theory 
compactification might look unnatural as the compactification 
limit of M-theory $\mu$- term in eqn. (\ref{akliopa}) involves only 
the T - moduli. 
However, at the large T limit, and assuming that the contribution of the 
auxiliary field U to supersymmetry breaking is very small, the 
 contributions of the T, U moduli are 
expected to be decoupled so that we can safely use the value of
(\ref{asdfgh}) in (\ref{akliopa}). In this case the K\"ahler potential
in eqn. (\ref{koilopi}) may be safely completed by the addition of 
the $-\ln(U + {\bar U})$ term.
In addition because of the special nature of solutions for the K\"ahler
potential \cite{owa} we must take the $B \rightarrow C$ limit in 
(\ref{asdfgh}). 
When the last considerations are taken into account the solutions
for the $\mu$-terms in the previous section may be lifted to its   
M-theory counterparts. 
If the T-moduli prove to be associated only to four dimensional $N=2$
 orbifold moduli then the B-terms correspond to $Z_2 
\times Z_4$ and $Z_2 \times Z_6$ $N=1$ orbifolds.
\newline
The 
$\mu$-term dependence of the B-term in (\ref{akliopa}) 
can be expressed in a general form in terms of $F^T$ and $F^S$
as  
\begin{eqnarray}
 \left(-3{\tilde C} cos\theta e^{-ia_S} -\sqrt{3}{\tilde C} 
sin\theta
+\frac{6{\tilde C} cos\theta (S + {\bar S})}{3(S+{\bar S})+ \alpha 
(T + {\bar T})}
+ \frac{2\sqrt{3}{\tilde C} sin\theta \alpha (T + {\bar T})}
{3 (S + {\bar S}) +
\alpha (T + {\bar T})} -1 \right) &\nonumber\\
+\;\;F^S (\frac{3}{2b}) + \frac{1}{2} (\frac{1}{(S + {\bar S})^2
m_{3/2}}) F_S^2 
 + \frac{1}{2} ( \frac{1}{(S + {\bar S})^2 m_{3/2}} )F_T^2 + F_T m_{3/2}
e^{\frac{3S}{2b}} \frac{\partial_T  \check{\mu} (T,U)}{\mu} ,  
\label{akliopaa1}
\end{eqnarray}
where we have parametrized the $\mu$-term solutions of the previous section
as
\begin{equation}
\mu=e^{G/2} e^{ \frac{3S}{2b} }  \check{\mu}(T,U) = e^{G/2}e^{\frac{3S}{2b}}
{\cal W}_{BC}
\label{akliao1}
\end{equation}
and have assumed (\ref{akliopa1}) and the following parametrization of the
dilaton and T-moduli auxiliary fields
\begin{equation}
F^S=e^{G/2} G_{{\bar S} S}^{-1} G_{\bar S},\;\;\;\;\;F^T=e^{G/2}G_{{\bar T} T}^{-1}
G_{\bar T}.
\label{ejkkso}
\end{equation}
For example for the M-theory compactifications flowing to the non-decomposable 
$Z_8$ -IIa $N=1$ orbifold the previous discussion applies with
\begin{eqnarray}
{\check{\mu}}(T,U)= - \frac{ \eta^{\prime}(T) }{\eta(T)}
[-\frac{1}{2}\eta^{-2}(\frac{T}{2})\eta^{-2}(U)](\partial_T 
\log\eta^2}{\frac{T}{2})(\partial_U \log\eta^2(U) ) \nonumber\\
-\frac{1}{2}[\eta^{-2}(\frac{T}{2}) \eta^{-2}(U)]
(\partial^2_T \log \eta^2(\frac{T}{2}))(\partial_U \log^2 \eta(U))
\nonumber\\
+ 2 \frac{\eta^{\prime}(T)}
{\eta(T)}[\eta^{-2}(T)\eta^{-2}(2U)](\partial_T \log\eta^2(T))(\partial_U 
\log \eta^2(2U))\nonumber\\ 
- [\eta^{-2}(T) \eta^{-2}(2U)](\partial_T^2 \log \eta^2(T))
(\partial_U \log \eta^2(2U)). 
\label{aklioas2}
\end{eqnarray}

\section{Conclusions}
In the context of heterotic string theory we calculated
4-dimensional non-perturbative superpotentials $\cal W$.
The dependence of the vector T, U moduli
arises 
from integrating out the massive compactification modes by summing over
their modular orbits.
Our calculation used the relation between the topological free 
energy and
non-perturbative superpotentials \cite{fklz} to perform this task.
The inclusion of the non-perturbative factor $e^{S}$ for the dilaton
was performed either from using the gaugino 
condensation \cite{veya,tay,lutay,fmtv}
in the hidden $E_8$ sector or using the BPS formula.  
In the literature \cite{lust} the same 
calculation that utilizes the sum over the perturbative  
modular orbits 
have been performed for the orbifold models 
invariant under the target spece duality group $SL(2,Z)_T \times
SL(2,Z)_U$.  We extended the previous result, by applying the 
technique of performing the sum over modular orbits 
of the chiral compactification modes,
to the more general classes of 
non-decomposable $N=1$ orbifolds. The latter 
string vacua exhibit target space duality groups
which are subroups of the modular
group $PSL(2, Z)$.
These solutions were later used to calculate solutions to the 
$\mu$-term problem.
The form of the $\mu$ term that we have proposed can be used 
to test observable CP violation effects in non-decomposable
orbifold compactifications of the heterotic string
in the spirit suggested in \cite{dugan,ibalu}.
The form of the $\mu$-terms associated with the non-decomposable 
orbifolds was used to examine the value of the B-soft parameter term 
obtained from compactifications of M-theory to four dimensions.  
Here we imposethe condition that at the weekly coupled limit
the B-theory M-terms flow to their 4D orbifold counterparts.


\begin{thebibliography}{700}  
\bibitem{dsww}M. Dine, N. Seiberg, X.-G.Wen and E. Witten, Nucl. Phys.
B278 (1986) 769;  Nucl. Phys. B289 (1987) 319. 
\bibitem{dise12}M. Dine and N. Seiberg, Phys. Rev. Lett. 21 (1986)
2625
\bibitem{kini}J. E. Kim and H. P. Nilles, Mod. Phys. Lett. A9 (1994)
3575; E. J. Chun, J. E. Kim and H. P. Nilles, Nucl. Phys. B370 (1992)
\bibitem{culu}G. Curio and D. Lust, Int. J. Mod. Phys. A12 (1997) 5847.
\bibitem{witta}R. Donagi, A. Grassi and E. Witten, Mod. Phys. Lett. A11
(1996) 2199.
\bibitem{casa}J. A. Casas and C. Munoz, Phys. Lett. B306 (1993) 288.
\bibitem{gima}G. F. Giudice and A. Masiero, Phys. Lett. B206 (1988) 480
\bibitem{BD1}T. Banks and L. Dixon, Nucl. Phys. B307 (1988) 93.
\bibitem{ANTO}I. Antoniadis, E. Gava, K. S. Narain, T. R. Taylor, Nucl. Phys.
B432 (1994) 187
\bibitem{dkl1}L. Dixon,V. Kaplunovsky and J. Louis, Nucl. Phys. B329 (1990) 27. 
\bibitem{kj}V. Kaplunovsky and J. Louis, Nucl. Phys. B444 (1995) 501.
\bibitem{deko}P. Mayr and S. Stieberger, Nucl. Phys. B407 (1993) 725;
D. Bailin, A. Love,\\ W. A. Sabra and S. Thomas, 
Mod. Phys. Letters. A9 (1994) 67; A10 (1995) 337 .
\bibitem{deko1}P. Mayr and S. Stieberger, Nucl. Phys. B407 (1993) 725;
\bibitem{erkl}J. Erler and A. Klemm, Commun. Math. Phys. 153 (1993) 579. 
\bibitem{oova}H. Ooguri and C. Vafa, Nucl. Phys. B361 (1991) 469. 
\bibitem{kiko}E. Kiritsis and C. Kounnas, Nucl. Phys. B503 (1997) 117
\bibitem{mastiebe}P. Mayr and S. Stieberger,  Phys. Lett. B355 (1995) 107
\bibitem{sp1}S.  Ferrara, C. Kounnas, M. Porrati and F. Zwigner, 
Nucl. Phys. B318 (1989) 75
\bibitem{sp2}S. Ferrara, C. Kounnas and F. Zwigner, Nucl. Phys. B429
(1994)
589, Erratum B433 (1995) 255 
\bibitem{fklz}S. Ferrara, C. Kounnas, D. L\"{u}st and F. Zwigner,
Nucl. Phys. B365 (1991)431.
\bibitem{dkl2}L. Dixon, V. Kaplunovsky and J. Louis,
Nucl. Phys. B{355} (1991) 649.
\bibitem{sen}J. Schwarz and A. Sen, Phys. Lett. B 312 (1993) 105;
A. Sen, Int. Jou. Mod. Phys. A 9 (1994) 3707
\bibitem{ropal}G. L. Cardoso, D. Lust and T. Mohaupt, Nucl. Phys. 
B455 (1995) 131. 
\bibitem{dhvw}L. Dixon, J. Harvey, C. Vafa and E. Witten,
Nucl. Phys. B261 (1985) 678; Nucl. Phys. B274 (1986) 285. 
\bibitem{imnq}L. E. Ib\'a\~nez, J.Mas, H. P. Nilles and F. Quevedo,
 Nucl. Phys. B301 (1988) 157.
\bibitem{kokos0}C. Kokorelis, Ph. D Thesis, Theoretical and 
Phenomenological Aspects\\ of Superstring 
Theories, Sussex, December 1997;
C. Kokorelis, String Loop Threshold Corrections from
Generalized Coxeter Orbifolds, SUSX-TH-98-010, to appear.
\bibitem{dfkz}J. P. Derendinger, S. Ferrara, C. Kounnas and F. Zwigner,
Nucl. Phys. B372 (1992) 145-188.
\bibitem{caf} A. Ceresole, R. D' Auria, S. Ferrara and A. Van Proyen,
Nucl. Phys. B444 (1995) 92  
\bibitem{wkll}B. de Wit, V. Kaplunovsky, J. Louis and D. Lust,
Nucl. Phys. B 451 (1995) 53
\bibitem{afgnt}I. Antoniadis, S. Ferrara, E. Gava, K. S. Narain and T. 
R. Taylor, Nucl. Phys. B 447 (1995) 35
\bibitem{kok}C. Kokorelis, Nucl. Phys. B542 (1999) 89
\bibitem{fosti} K. Foerger and S. Stieberger, Nucl. Phys. B514 (1998) 135   
\bibitem{bim}A. Brignole, L. I. Ibanez and C. Munoz,
Nucl. Phys. B422: (1995) 125, Erratum : Nucl. Phys. B476, (1995) 747. 
\bibitem{sce}A. Brignole, L. E. Ibanez, C. Munoz and C. Scheich, 
Z. Phys. C74 (1997) 157.
\bibitem{bafs}V. Kaplunovsky and J. Louis,  Phys. Lett. B306 (1993) 269.
\bibitem{blst1}D. Bailin, A. Love, W. Sabra and S. Thomas;
Phys. Lett. B320 (1994) 21.
\bibitem{blst2}D. Bailin, A. Love, W. Sabra and S. Thomas,
Mod. Phys. Lett. (1994) 1229.
\bibitem{clm1}G. L. Cardoso, D. Lust and T. Mohaupt, Nucl. Phys. B432 
(1994) 68.
\bibitem{lust}G. L. Cardoso, D. Lust and T. Mohaupt, Nucl. Phys. B450
(1995) 115.
\bibitem{dugan}M. Dugan, B. Grinstein and L. Hall, Nucl. Phys. B255,
(1985) 413.
\bibitem{ibalu}L. E. Ibanez and D. Lust, Phys. Lett. B267 (1991) 51 
\bibitem{Love2}B. Acharya, D. Bailin, A. Love, W. A. Sabra and S.
Thomas,
Spontaneous breaking\\ of CP by Orbifold Moduli, Phys. Lett. B357 (1995)
387; D. Bailin,\\ G. V. Kraniotis  and A. Love,
 Supersymmetric CP problem in Orbifold \\Compactifications,
Nucl. Phys. B518 (1998) 92.  
\bibitem{howi}P. Horava and E. Witten, Nucl. Phys. B460 (1996) 506;
Nucl. Phys. B475 (1996) 94. 
\bibitem{chkimi}K. Choi, H. B. Kim and C. Munoz, Phys. Rev. D57 
(1998) 7521.
\bibitem{nils}H. P. Nilles, M. OLechowski and M. Yamagichi,
Phys. Lett. B145 (1997) 24; Nucl. Phys. B530 (1998) 43.
\bibitem{owa}A. Lukas, B. A. Ovrut and D. Waldram, Phys. Rev. D57
(1998) 7529; A. Lukas, B. A. Ovrut and D. Waldram, 
Nucl. Phys. B532 (1998) 43.  
\bibitem{bakralo}D. Bailin, G. V. Kraniotis and A. Love, 
Phys. Lett. B432 (1998) 323, Sparticle Spectrum and Dark Matter in 
M-theory.
\bibitem{nano}T. Li, J. L. Lopez and D. V. Nanopoulos,
Mod. Phys. Lett. A12 (1997) 2647;Phys. Rev. D 56 (1997) 2602. 
\bibitem{li}T. Li, Phys. Rev. D 57 (1988) 7539.
\bibitem{latho}Z. Lalak and S. Thomas, Nucl. Phys. B515 (1988) 55.
\bibitem{anqu}I. Antoniadis and M. Quiros, Phys. Lett. B392 
(1997) 61.
\bibitem{cacamu}B. de Carlos, J. A. Casas and C. Munoz, Nucl. Phys. B 399
(1993) 623.
\bibitem{homou}J. Horne and G. Moore, Nucl. Phys. B432 (1994) 109.  
\bibitem{ducro}E. Dudas and C. Crojean, Nucl. Phys. B507 (1997) 553
\bibitem{veya}G. Veneziano and S. Yankielowicz, Phys. Lett. B113  (1982) 231.
\bibitem{tay}T. R. Taylor, Phys. Lett. B252 (1990) 59.
\bibitem{lutay}D. Lust and T. Taylor, Phys. Lett. B253 (1991) 335
\bibitem{fmtv}S. Ferrara, N Magnoli, T. R. Taylor and G. Veneziano,
Phys. Lett. B245 (1990) 409.  
\end{thebibliography}
\end{document}